\documentclass[12pt]{iopart}

\usepackage{graphics}
\usepackage{epsfig}
\usepackage{xspace}
\usepackage{amssymb}
\usepackage{color}

\newcommand {\tpsi}{\tilde{\psi}}

\newcommand{\PT}{{\cal PT}}

\newcommand{\cE}{{\cal E}}
\newcommand{\tx}{\tilde{x}}

\newcommand{\tOmega}{\tilde{\Omega}}
\newcommand{\tilt}{\tilde{t}}
\newcommand{\cL}{\cal{L}}

\newcommand{\be}{\begin{equation}}
\newcommand{\ee}{\end{equation}}

\newcommand{\sech}{\mathrm{sech}}

\begin{document}

\title[Solitons in a Hamiltonian  $\PT$-symmetric coupler]{Solitons in a Hamiltonian  $\PT$-symmetric coupler}

\author{Dmitry A. Zezyulin$^{1,2}$ and Vladimir V. Konotop$^3$}

\address{$^1$ITMO University, St. Petersburg 197101, Russia \\
	$^2$Institute of Mathematics with Computer Center, Ufa Scientific Center, Russian Academy of Sciences, Chernyshevskii str., 112, Ufa  450008, Russia\\
	$^3$Centro de F\'{\i}sica Te\'orica e Computacional and Departamento de F\'{\i}sica, Faculdade de Ci\^encias,
	Universidade de Lisboa,
	Campo Grande, Ed. C8, Lisboa  1749-016, Portugal}
\ead{dzezyulin@corp.ifmo.ru}

\vspace{10pt}

\begin{abstract}
We introduce a nonlinear parity-time-symmetric  dispersive 
coupler which admits Hamiltonian and Lagrangian formulations.  We show that,  in spite of the  gain and dissipation, the model has  several conservation laws. The system also supports a variety of exact solutions. We focus on    exact bright solitons and  demonstrate numerically that they are dynamically stable in a wide parameter range and undergo elastic interactions, thus manifesting nearly-integrable dynamics.  Physical applications of the introduced model in the theory of Bose-Einstein condensates in nonlinear lattices are discussed.
\end{abstract}

%
%
%
%
%

\section{Introduction}

The relation between Hermitian quantum mechanics and its parity-time ($\PT$--) symmetric extension became the focus of intense debates shortly after the latter paradigm was introduced in~\cite{Bender}. In particular, it was shown that any $\PT$-symmetric Hamiltonian with purely real spectrum, i.e., in the unbroken $\PT$-symmetric phase,  can be transformed  to a   Hermitian one using a similarity transformation 
\cite{Mostaf2002}. This fact   establishes certain equivalence between the two formulations. In the classical limit, $\PT$-symmetric models occupy an ``intermediate position'' between conservative and dissipative nonlinear systems, featuring properties of both these types (see the discussion in~\cite{KYZ}). A typical real-world $\PT$-symmetric system consists of an active element coupled to a lossy one, {\color{black} like those theoretically introduced~\cite{ElGan07} and experimentally studied~\cite{Ruter} in the non-Hermitian discrete optics.} Generally, gain and dissipation break the conservation laws for energy or  other physically relevant quantities. This strongly affects the nonlinear dynamics,  making impossible (or highly nontrivial) its description using the analytical approaches, available, say, for Hamiltonian systems.

Recently, it has been discovered  that dynamics of some nonlinear $\PT$-symmetric models with gain and dissipation still can be described using the Hamiltonian formalism  
and is therefore  characterized by a rather  high degree of regularity. More specifically, the Hamiltonian structure has been revealed for  $\PT$-symmetric  coupled oscillators~\cite{coupled_oscil}, for some  completely  integrable dimer models~\cite{Ramezani,BG14,Barash_Hamilt,	JC93,JCAH93,BPD15}, and  for  chains of $\PT$-symmetric pendula \cite{CP}. We also mention that systems  which do not posses $\PT$ symmetry but display characteristics of conservative and dissipative ones, are also known; they are described by time-reversible Hamiltonians~\cite{PoOpBa}. However, all the mentioned examples belong to the class of  dynamical systems, i.e., the corresponding models are described by  systems of  ordinary differential equations.  

The main goal of this paper is to introduce a nonlinear  $\PT$-symmetric  dispersive system   which incorporates  gain and losses and at the same time allows for the Hamiltonian formulation. This  model  emerges as a $\PT$-generalization of the resonant four-wave mixing process in a spinor Bose-Einstein condensate  loaded in linear and nonlinear lattices.  We demonstrate that such a two-component model, where one component experiences gain and another one loses   atoms, can be obtained from a real-valued Hamiltonian. Next, we explore the associated conserved quantities and present various exact solutions and some traits of the system's nonlinear dynamics. We in particular  focus on exact bright soliton solutions and demonstrate numerically that they are   stable in a wide parameter range.

The paper is organized as follows. In Sec.~\ref{sec:model} we discuss a physical example where the proposed $\PT$-symmetric model  arises. The model itself and its basic properties are presented in Sec.~\ref{sec:main}. Section~\ref{sec:exact} demonstrates that the introduced system admits a variety of exact solutions which can be written down in the analytical form, and Sec.~\ref{sec:solitons} studies exact solutions in the form of bright solitons. Finally, Sec.~\ref{sec:concl} concludes the paper and outlines some perspectives.

\section{On the physical model}
\label{sec:model}

{\color{black} To introduce the model, we start with a physical example.  We  consider a one-dimensional Bose-Einstein condensate (BEC) 
	loaded in a linear lattice  which experimentally can be created by a two counter-propagating laser beams~\cite{lin_latt}.   We also assume that the scattering length varies periodically in space, i.e. creates a nonlinear lattice.  The latter can be created, for example, by periodically varying external field affecting the scattering length by means of the Feshbach resonance.  Nonlinear lattices have been realized in laboratory~\cite{nonlin_latt} and studied in numerous theoretical works (see e.g. Refs.~\cite{NL_meanfiled,Trippenbach}). In particular, in \cite{Trippenbach} it was shown that matching condition for resonant four-wave processes can be achieved by modification the momentum,  while in the linear lattices the matching condition is achieved by the modifying the dispersion relation ~\cite{lin_matching}.}

{\color{black}
	The meanfield Hamiltonian describing the BEC in linear and nonlinear lattices reads
	\begin{eqnarray}
	\label{Hmilt_BEC}
	\hat{H}_{\rm BEC}=\int_{-\infty}^{\infty}\left\{  {\Psi}^*\hat{h}{\Psi}+\tOmega \cos(\nu t)V_1(x) |\Psi|^2 +  \chi(x) [1+2\cos(2\nu t)]    |\Psi|^4  \right\} dx,
	\end{eqnarray} 
	where $\Psi$ is an order parameter $\hat{h}=-\partial_x^2  +V_{0}(x)$, $V_{0,1}(x)=V_{0,1}(x+L)$ are the even ("$0$") and odd ("$1$")  components of the optical lattice:  $V_0(x)=V_0(-x)$, $V_1(x)$ and $V_1(x)=-V_1(-x)$, $L$ is the lattice period, $\tOmega\ll 1$ is the small parameter defining relative depth of the odd component, and $\chi(x)=\chi(x+L)$ describes the nonlinear lattice which undergoes periodic oscillations with the frequency $2\nu$. The value of $\nu$  will be specified below. The amplitude of the shallow odd lattice $V_1(x)$ also periodically varies in time but with the frequency $\nu$. In (\ref{Hmilt_BEC}) dimensionless units with $\hbar=1$ and $m=1/2$, where $m$ is the atomic mass, are used.
}

\begin{figure}
	\begin{center}
		\includegraphics[width=0.8\textwidth]{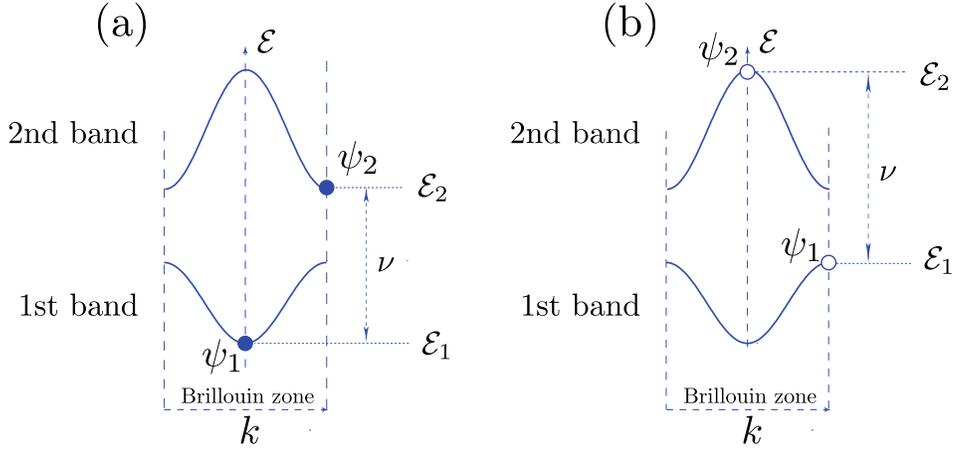}
	\end{center}
	\caption{{\color{black}  Two schematic representations of  resonantly interacting pairs of Bloch modes in the linear lattice. Panels (a) and (b) correspond to two pairs of modes (shown with circles) with positive and negative curvatures of dispersion relations (i.e.,  effective masses), respectively. In both panels, $k$ is the Bloch momentum.}}
	\label{fig:bands}
\end{figure}

{\color{black}  Let us consider  the evolution of a superposition $\Psi=\Psi_1(x,t)+\Psi_{2}(x,t)$ of two wave packets of Bloch states. Choosing these states as shown in Fig.~\ref{fig:bands}, i.e. Bloch modes with zero group velocities and having energies at the  band edges  with equal signs of   curvatures of the dispersion relations, one can look for a solution in the form 
	\begin{eqnarray}
	\label{ansatz}
	\begin{array}{l}
	\displaystyle{	\Psi_1(x,t)=\left [\epsilon u(\tx,\tilt)\psi_1(x)+\epsilon^2 \frac{\partial u}{\partial\tx }\tpsi_1(x) +\cdots\right] e^{-i\cE_1 t},
		}
	\\[3mm]
	\displaystyle{
	\Psi_2(x,t)=\left[\epsilon v(\tx,\tilt)\psi_2(x)+\epsilon^2 \frac{\partial v}{\partial\tx }\tpsi_2(x) +\cdots\right]e^{-i\cE_2 t}.
}
	\end{array}
	\end{eqnarray}
	Here $\epsilon\ll 1$ is a formal small parameter, $u(\tx,\tilt)$ and $v(\tx,\tilt)$ are functions of   slow variables $\tx=\epsilon x$ and $\tilt=\epsilon^2 t$, i.e., envelopes of the Bloch states $\psi_{1,2}$ corresponding to the energies $\cE_{1,2}$: $\hat{h}\psi_j=\cE_j\psi_j$ ($j=1,2$). 	In this case, $\psi_{1}$ and $\psi_{2}$ are real-valued periodic functions, as it follows from the Floquet theorem). Detailed calculations of other functional coefficients of the expansions (\ref{ansatz}), i.e., functions $\tpsi_{1,2}(x)$ can be performed within the framework of the multiple scale analysis. Here we do not present all the details as they are available in the literature (see e.g.~\cite{KS}), but only show their effect on the Hamiltonian structure of the model. In the meantime, it will be important that $\psi_j$ and $\tpsi_j$ are orthogonal to each other, which in our case means that $ \int_{-L/2}^{L/2}\psi_j(x)\tpsi_j(x)dx=0$. 
	Thus the dynamics of the superposition $\Psi$ is completely determined by the evolution of $u(\tx,\tilt)$ and $v(\tx,\tilt)$, and our aim is to obtain the Hamiltonian of this dynamics.
}

{\color{black} Now we use   substitution (\ref{ansatz}) in the Hamiltonian (\ref{Hmilt_BEC}) and require $\nu$ to be the difference between the energies of the states: $\nu=\cE_2-\cE_1$ (see Fig.~\ref{fig:bands}).  The formulated assumptions allow   to approximate different terms as follows:
	\begin{eqnarray}
	\label{aux1}
	\int_{-\infty}^{\infty}{\Psi}^*\hat{h}{\Psi}dx
	\approx \epsilon^2   \int_{-\infty}^{\infty}\left(\cE_1 \psi_1^2|u|^2+\cE_2 \psi_2^2|v|^2\right)dx
	\nonumber \\
	+
	\epsilon^4\int_{-\infty}^{\infty} 
	 \left(\psi_1^2-\tpsi_1\frac{d\psi_1}{dx}-\tpsi_1\hat{h}\tpsi_1\right)|u_{\tx}|^2dx  	\nonumber\\	
	 +	\epsilon^4\int_{-\infty}^{\infty} 
	\left(\psi_2^2-\tpsi_2\frac{d\psi_2}{dx}-\tpsi_2\hat{h}\tpsi_2\right)|v_{\tx}|^2 dx, 
	\end{eqnarray}
	where we dropped the terms $\sim \psi_jd\psi_j/dx$ whose contribution to the integral is negligible due to the opposite parities of $\psi_j$ and  $d\psi_j/dx$, as well as the terms $\sim \psi_j \tpsi_j$  due to their orthogonality;
}
{\color{black}	
	\begin{eqnarray}
	\label{aux2}
	\int_{-\infty}^{\infty} 2\cos(\nu t)V_1(x) |\Psi|^2dx\approx 
	\epsilon^2   \int_{-\infty}^{\infty}V_1(x) \psi_{1}\psi_2 \left(u^*v+uv^*\right)dx
	\end{eqnarray} 
	where we neglected the terms $\sim V_1(x)|\psi_{j}|^2$ because of their symmetry (these terms can also be discarded in the rotating wave approximation as rapidly varying   $\sim \cos (\pm\nu t)$);
	\begin{eqnarray}
	\label{aux3}
	\int_{-\infty}^{\infty}\chi(x) |\Psi|^4dx\approx  \epsilon^4  \int_{-\infty}^{\infty}\chi(x)\left( \psi_{1}^4|u|^4 + \psi_{2}^4|v|^4+ 4\psi_{1}^2\psi_{2}^2|u|^2|v|^2\right)dx
	\end{eqnarray}
	where the terms oscillating as $e^{\pm i\nu t}$ and $e^{\pm 2 i\nu t}$
	are dropped as rapidly oscillating;
	\begin{eqnarray}
	\label{aux4}
	\int_{-\infty}^{\infty}2\cos(2\nu t) \chi(x)|\Psi|^4dx\approx  \epsilon^4  \int_{-\infty}^{\infty} \chi(x)\psi_{1}^2\psi_{2}^2\left[   u^2(v^*)^2+  (u^*)^2 v^2  \right]dx
	\end{eqnarray}
	where all other terms are rapidly oscillating. 
} 

\textcolor{black}{The term $\sim \epsilon^2$ in   (\ref{aux1}) amounts to $\int_{-\infty}^{\infty} (\cE_1 |\Psi_1|^2 + \cE_2 |\Psi_2|^2)dx$. It   does not affect the dynamics of slowly varying amplitudes $u$ and $v$, but simply describes   the leading order of the Hamiltonian equation
	\begin{equation}
	\label{Hmilt}
	i\frac{\partial \Psi}{\partial t} = \hat{H}_{BEC}\Psi,
	\end{equation} 
	whose left hand side reads
	\begin{equation}
	\label{time}
	i\frac{\partial \Psi}{\partial t} = \cE_1 \Psi_1 + \cE_2 \Psi_2 + i\epsilon^3\left(e^{-i\cE_1 t}\psi_1\frac{\partial u}{\partial \tilde{t}} + e^{-i\cE_2 t}\psi_2\frac{\partial v}{\partial \tilde{t}} \right) 
	\end{equation} }

{\color{black} For other  integrals in the right hand sides of (\ref{aux1})-(\ref{aux4}), we note that they contain functions depending on  fast ($\psi_{1,2}$) and slow ($u,v$)  variables. In order to obtain formally the effective Hamiltonian for   slow envelopes $u$ and $v$, we substitute the fast functions in the integrands by their mean values over the lattice period $L$. Then assuming the Bloch functions normalized, we obtain 
	$\displaystyle{\psi_{j}^2\to
		\int_{-L/2}^{L/2}\psi_{j}^2(x)dx=1}$. Next,   
	we assume that the linear lattice is chosen such that 
}
{\color{black}
	
	\begin{eqnarray}
	\int_{-L/2}^{L/2}\left(\psi_1^2-\tpsi_1\frac{d\psi_1}{dx}-\tpsi_1\hat{h}\tpsi_1\right) dx=
	\int_{-L/2}^{L/2}\left(\psi_2^2-\tpsi_2\frac{d\psi_2}{dx}-\tpsi_2\hat{h}\tpsi_2\right) dx
	\nonumber \\
	:=\frac{1}{2m_{eff}}
	\end{eqnarray}
	were $m_{eff}$ is the effective mass which is proportional to the radius of curvature of the dispersion curves~\cite{KS} and can be positive as in Fig.~\ref{fig:bands}(a) or negative as in Fig.~\ref{fig:bands}(b). To simplify the  consideration, we address only the case of positive effective mass: $m_{eff}>0$ (the generalization for $m_{eff} <0$ is straightforward).
} 
{\color{black}
Additionally, we assume that   one can find a constant $g$ such that  within the accepted accuracy 
	\begin{eqnarray}
	\label{cond}
	\frac{g}{2}= \int_{-L/2}^{L/2} \chi(x)\psi_{1}^2\psi_{2}^2dx=\int_{-L/2}^{L/2}\chi(x) \psi_{1}^4dx = \int_{-L/2}^{L/2} \chi(x)\psi_{2}^4dx.  
	\end{eqnarray}
	Finally, we consider weak coupling, allowing to define $\Omega = {\cal O}(1)$ through the relation 
	\begin{equation}
	\epsilon^2\Omega = \tOmega\int_{-L/2}^{L/2} V_1(x)\psi_{1}\psi_2 dx
	\end{equation}
} 	

{\color{black}
	Having substituted the products of the fast functions  by their average values, the integrals of the  remaining slow envelopes can  be computed with respect to the renormalized slow variable $\tx$ using that   $$dx  = 
	\frac{1}{\epsilon} d\tx.$$ Then collecting all the integrals  $\sim \epsilon^3$ with respect to $d\tx$   leads to the energy functional 
	\begin{eqnarray}
	\label{energy}
	E= \int_{-\infty}^\infty \Bigl[|u_x|^2 + |v_x|^2 + \Omega (uv^* + u^* v)  \nonumber \\
	+  \frac{g}{2}(|u|^4 + |v|^4+u^2(v^*)^2+(u^*)^2 v^2 + 4|u|^2|v|^2)\Bigr]dx.
	\end{eqnarray}
	where we have omitted  tildes over $x$, since the rest of the paper we deal  only with the envelopes $u$ and $v$, and have renormalized the coefficients  $2 m_{eff} \Omega \to \Omega$, $2  m_{eff} g  \to g$. }

The equations for the evolution of the slow amplitudes are obtained from {\color{black} Schr\"odinger equation (\ref{Hmilt}) with the time derivative given by (\ref{time}), projected over $\psi_1$ and $\psi_2$, and can be expressed in the Hamiltonian form}
\begin{eqnarray}
\label{eq:motionE}
\frac{\delta E}{\delta u} = -iu_t^*, \quad \frac{\delta E}{\delta v} = -iv_t^*, \quad
\frac{\delta E}{\delta u^*} =  iu_t, \quad
\frac{\delta E}{\delta v^*} =  iv_t,
\end{eqnarray}
which gives
\begin{equation}
\label{eq:nonlin1D_cons}
\eqalign{iu_t = -u_{xx} +\Omega v + g |u|^2 u + 2g |v|^2u + g u^*v^2,\\
	iv_t = -v_{xx}  +\Omega u + g |v|^2 v+  2g |u|^2v + g  u^2v^*.}
\end{equation}

For  $x$-independent solutions system (\ref{eq:nonlin1D_cons})  reduces to that  explored in~\cite{Trippenbach}.  In general, in the basis of new functions $u_{\pm}(x, t) = u(x,t) \pm v(x,t)$ system (\ref{eq:nonlin1D_cons}) decouples into two nonlinear Schr\"odinger equations \cite{Gergjikov,IB}
\begin{equation}
\label{eq:IB}
iu_{\pm,t} = -u_{\pm, xx} \pm \Omega u_\pm  + g |u_\pm|^2 u_\pm,
\end{equation}
and is therefore completely integrable. 
Another interesting observation is  that the obtained system  (\ref{eq:nonlin1D_cons})  has a {\em bi-Hamiltonian}~\cite{AC91} structure. Indeed, it can also be obtained from the Hamiltonian 
\begin{equation}
\label{eq:Hlin_cons}
H = \int_{-\infty}^\infty \left[ u_x^*  v_x +  u_x    v_x^* + \Omega (|u|^2 + |v|^2)   +  g(|u|^2+|v|^2)(u^*v + uv^*) \right] dx
\end{equation}
in the new canonical variables  according to the following equations:
\begin{eqnarray}
\label{eq:motion}
\frac{\delta H}{\delta u} = -iv_t^*, \quad \frac{\delta H}{\delta v} = -iu_t^*,\quad 
\frac{\delta H}{\delta u^*} =  iv_t, \quad
\frac{\delta H}{\delta v^*} =  iu_t.
\end{eqnarray}

\section{Hamiltonian $\PT$-symmetric coupler}
\label{sec:main}

Hamiltonian equations (\ref{eq:motion}) feature  the \textit{cross-gradient} structure which has been recently revealed for some  $\PT$-symmetric Hamiltonian systems of coupled oscillators  \cite{coupled_oscil,BG14,BPD15}. This observation   opens the route to generalize  the system (\ref{eq:nonlin1D_cons}) by including $\PT$-symmetric gain and loss terms. To this end, let us introduce the following  generalization of the  Hamiltonian (\ref{eq:Hlin_cons}):
\begin{eqnarray}
\label{eq:Hlin_PT}
H = \int_{-\infty}^\infty \left [u_x^*  v_x +  u_x    v_x^* + i\gamma (uv^* - u^*v) + \Omega (|u|^2 + |v|^2) \right.
\nonumber \\
\hspace{4cm}\left.   +  g(|u|^2+|v|^2)(e^{i\phi}u^*v + e^{-i\phi}uv^*) \right] dx,
\end{eqnarray}
where $\phi$ is a real constant.
Then,   applying     equations  (\ref{eq:motion}) to  the Hamiltonian (\ref{eq:Hlin_PT}), we arrive at the   \textit{Hamiltonian $\PT$-symmetric  coupler}:
 \begin{equation}
 \label{eq:nonlin1D}
 \eqalign{
 iu_t = -u_{xx} + i\gamma u +\Omega v + g e^{-i\phi}  |u|^2 u + 2g e^{-i\phi}  |v|^2u + ge^{i\phi} u^*v^2,\\
 iv_t = -v_{xx} - i\gamma v +\Omega u + ge^{i\phi} |v|^2 v+  2g e^{i\phi} |u|^2v + ge^{-i\phi}  u^2v^*.}
 \end{equation}
 \textcolor{black}{To the best of our knowledge, system (\ref{eq:nonlin1D}) has not been considered in the previous literature. This system  will be in the focus of our attention in the rest of this study.}
From the physical perspective,    system (\ref{eq:nonlin1D})   can be considered as the described above  model of the spinor BEC in the nonlinear lattice, where atoms are loaded in the $u$-component and are eliminated from the $v$-component with the strength characterized by the gain-loss coefficient $\gamma$.  The model (\ref{eq:nonlin1D}) also includes  nonlinear gain and losses due to inelastic two-body interactions characterized by the real parameter $\phi$. Without loss of generality, in what follows we assume that $\gamma, \Omega \geq 0$.
 
 {\color{black} Model (\ref{eq:nonlin1D}) belongs to the class of nonlinear dispersive systems. Indeed, considering the propagation of small-amplitude plane waves $(u,v) = (p, q) e^{i(kx-\omega t)}$,  where $|p|, |q| \ll 1$,   $\omega$ is the frequency, and $k$ is the wavenumber,  we obtain  the two branches  of dispersion relation  in the form
\begin{equation}
\label{eq:disp}
\omega_\pm(k)=k^2\pm\sqrt{\Omega^2-\gamma^2}.
 \end{equation}
 Thus the waves propagating in such a coupler have  dispersive nature and, in order to reflect this fact, the system (\ref{eq:nonlin1D}) can be referred to as a dispersive nonlinear $\PT$-symmetric coupler.}
 
 {\color{black}  To complete the formulation of the model, we note that underlying  physical setup   for system  (\ref{eq:nonlin1D}) resulted from  the process of four-wave mixing in  linear and nonlinear lattices modifying, respectively, energy and momentum conservation laws. It turns out that the presence of gain and loss in (\ref{eq:nonlin1D}) allows one to modify the dispersion relation, which allows for obtaining matching conditions for four-wave mixing in terms of the two-component field $(u,v)$~\cite{Marek}. Thus the phenomenon of the wave mixing is also expectable  in the framework of the model (\ref{eq:nonlin1D}). The peculiarities of this effect are beyond the scope of the present work: here we concentrate mainly on localized solitonic solutions.}

In the linear case ($g=0$)    system (\ref{eq:nonlin1D}) assumes the form 
\begin{equation}
	i\frac{\partial\ }{\partial t} \left(\begin{array}{c}
u\\v
	\end{array}\right) = {L}  \left(\begin{array}{c}
	u\\v
	\end{array}\right), \quad  {L} = \left(\begin{array}{cc}
	-\partial^2_x + i\gamma &\Omega  \\\Omega & -\partial^2_x - i\gamma
	\end{array}\right).
\end{equation}
Linear operator $L$ is $\PT$ symmetric:   it commutes with the $\PT$ operator,  where ${\cal P}$ swaps the components,
\begin{equation}
{\cal P} \left(\begin{array}{c} u\\v\end{array}\right) = \left(\begin{array}{c} v\\u\end{array}\right),
\end{equation}
and $\cal T$ is the  component-wise complex conjugation combined with the time reversal: ${\cal T }u(x,t) = u^*(x,-t)$, ${\cal T }v(x,t) = v^*(x,-t)$. \textcolor{black}{As readily follows from the dispersion relation (\ref{eq:disp})}, the linear  coupler is stable (i.e., $\PT$ symmetry is unbroken) if $\gamma/\Omega<1$.  For  $\gamma/\Omega\geq 1$ there are unbounded in time solutions (i.e., $\PT$ symmetry is broken). The transition from unbroken to broken $\PT$ symmetry, i.e., the situation $\gamma = \Omega$, corresponds to the exceptional point (EP).
 
The nonlinear ($g\ne 0$) system (\ref{eq:nonlin1D}) is $\PT$ symmetric in the following sense:  if functions $u(x,t)$ and $v(x,t)$ solve   equations    (\ref{eq:nonlin1D}) in some domain $(x,t)\in \mathbb{R}\times [-t_0, t_0]$, then the new functions $u_{\PT}(x,t):={v}^*(x,-t)$ and $v_{\PT}(x,t):={u}^*(x,-t)$ also solve system (\ref{eq:nonlin1D})  in the same domain.

The model (\ref{eq:nonlin1D}) generically 
does not conserve the number of particles ($L^2$- norm of the solution)
\begin{equation}
N=\| u \|_{L^2}^2 + \| v \|_{L^2}^2  = \int_{-\infty}^\infty (|u|^2+|v|^2)dx.
\end{equation}
Indeed, it is straightforward to obtain the relation
\begin{equation}
\label{eq:dN}
\frac{dN(t)}{dt} = 2\gamma\int_{-\infty}^\infty (|u|^2 - |v|^2)\ dx  - 2g\sin\phi\int_{-\infty}^\infty (|u|^4 - |v|^4)\ dx.
\end{equation}
At the same time, by construction, system (\ref{eq:nonlin1D}) conserves  the real-valued Hamiltonian $H$ given by (\ref{eq:Hlin_PT}): $dH/dt=0$. In order to identify other  conserved quantities, it is convenient to use  the Lagrangian formalism, starting with the (real-valued) Lagrangian density
\begin{eqnarray}
{\cal L} = \frac{i}{2}(u_tv^* - u_t^*v + u^*v_t - uv_t^*) - [ u_x^* v_x + u_xv_x^*+i\gamma(uv^*-u^*v)  \nonumber  \\ \hspace{2cm}
 + \Omega(|u|^2+|v|^2) +  g(|u|^2+|v|^2)(e^{i\phi}u^*v + e^{i\phi}uv^*)].
\end{eqnarray}
Then   system (\ref{eq:nonlin1D}) is equivalent to the Euler--Lagrange equations
\begin{equation}
{\frac{\partial \cL}{\partial u}} = \frac{\partial\ }{\partial x}  \left(\frac{\partial\cL}{\partial u_x}\right) + \frac{\partial\ }{\partial t} \left(\frac{\partial\cL}{\partial  u_t}\right),\qquad 
{\frac{\partial \cL}{\partial v}} = \frac{\partial\ }{\partial x}  \left(\frac{\partial\cL}{\partial v_x}\right) + \frac{\partial\ }{\partial t} \left(\frac{\partial\cL}{\partial  v_t}\right).
\end{equation}
Since the  action functional  $S=\int_0^t\int_{-\infty}^\infty {\cL}\, dx dt$ is invariant under  
space and time translations, as well as under the  phase rotation, from Noether's theorem (see e.g.~\cite{Sulem}) we obtain three conserved quantities for the model (\ref{eq:nonlin1D}): one of them corresponds to the Hamiltonian  $H$ in (\ref{eq:Hlin_PT}), and two other correspond to  the quasi-power
\begin{equation}
\label{eq:Q}
Q = \int (uv^*+u^*v)dx,\qquad \frac{dQ}{dt}=0,
\end{equation}
and the  quasi-momentum 
\begin{equation}
P = i\int_{-\infty}^\infty ( u_x v^* -  u_x^* v)dx, \qquad \frac{dP}{dt}=0.
\end{equation}

As an  immediate consequence of the established conservation laws, we observe that the   total number of particles   (as well as the $L^2(\mathbb{R})$-norm  of the solution) is bounded from below by a nonnegative  constant:
\begin{equation}
\label{eq:N}
N(t)   \geq |Q(t)|=|Q(0)|.
\end{equation}
Additionally,  the total $H^1$-norm
is also bounded from below by a nonnegative constant which is {\it a priori} defined  by the   quasi-momentum:
\begin{equation}
N(t) + \|u_x\|_2^2 + \|v_x\|_2^2  = \| u\|_{H^1}^2 +  \| v\|_{H^1}^2 \geq |P(t)|=|P(0)|.
\end{equation}
Here, we used the standard definition of $H^1(\mathbb{R})$-norm, that is
\begin{equation}
\| u\|_{H^1}^2 = \| u\|_{L^2}^2 +    \| u_x\|_{L^2}^2, \quad \| v\|_{H^1}^2 = \| v\|_{L^2}^2 +    \| v_x\|_{L^2}^2.
\end{equation}

\section{Reductions and exact solutions}
\label{sec:exact}

Introduced in the previous section $\PT$-symmetric Hamiltonian coupler   (\ref{eq:nonlin1D}) admits several important reductions and allows for exact analytical solutions in the form of continuous families of solitons. To obtain them, we look for solutions in the form 
\begin{eqnarray}
\left(\begin{array}{c}
u \\ v
\end{array}\right)=R_1R_2\left(\begin{array}{c}
U \\ V
\end{array}\right),
\end{eqnarray}
where $R_1$ and $R_2$ are rotation matrices
\begin{eqnarray}
R_1=\left(\begin{array}{cc}
e^{-i\delta/2} &-e^{i\delta/2}\\
e^{i\delta/2} &e^{-i\delta/2}
\end{array}\right) \qquad 
R_2=\left(\begin{array}{cc}
\cos(\alpha) & -\sin(\alpha)
\\
\sin(\alpha) & \cos(\alpha)
\end{array}\right)
\end{eqnarray}
and real parameters  $\delta$  and $\alpha$ are to be defined.
The resulting equations are not shown here as they appear too lengthy. However, it is straightforward to verify that for the certain choice of parameters they allow for ``one-component'' solutions with  $U\neq 0$ and $V\equiv0$ (or $U\equiv 0$ and $V\ne 0$).  {\color{black}  These ``one-component'' solutions correspond to the situation when components $u$ and $v$ are proportional, i.e., $u(x,t)=r v(x,t)$, where $r$ is a time- and space-independent constant.}

\subsection{Unbroken linear $\PT$ symmetry}

\label{sec:below}
In the domain of unbroken linear $\PT$ symmetry we have $\gamma\leq \Omega$. For the existence of a one-component solution  $U\neq0$ and $V\equiv0$ we  require   $\alpha=0$ and   $\sin(\delta) =- \gamma/\Omega$ \cite{DM11}, i.e.,
\begin{eqnarray}
\label{delta12}
\delta=\delta_1=-\arcsin (\gamma/\Omega) \quad \mbox{or}  \quad  \delta=\delta_2 = \pi +\arcsin (\gamma/\Omega).
\end{eqnarray}
Additionally, the coefficient $\phi$ depends on $\gamma$ as
\begin{equation}
\label{eq:phi}
\phi_{1,2} = -\arctan\left(\frac{\sin(2 \delta_{1,2})}{\cos(2\delta_{1,2})-3}\right) =  \mp \arctan\left(\frac{\gamma\Omega\sqrt{\Omega^2-\gamma^2}}{\Omega^2+\gamma^2}\right).
\end{equation}
Now the system is reduced to the standard  (conservative) nonlinear Schr\"odinger (NLS) equation with real coefficients
\begin{equation}
\label{eq:NLS}
iU_t = -U_{xx} \pm  \sqrt{\Omega^2-\gamma^2}U +  \frac{4g\Omega}{\sqrt{3\gamma^2 + \Omega}}|U|^2U,
\end{equation}
where  the upper and the lower signs correspond to $\delta_1$ and $\delta_2$  in (\ref{delta12}).  The original fields can be  recovered from $U$ as  $u=e^{-i\delta/2} U$ and $v=e^{i\delta/2} U$.

The NLS equation (\ref{eq:NLS})   is completely integrable and supports a variety of exact solutions, including bright ($g < 0$) and dark  ($g > 0$)  solitons \cite{AC91, Sulem}, rogue waves and breathers, etc. (see e.g. \cite{Peregrine(1983), AKM87}). Each of these solutions  has   two counterparts [for two choices of $\delta$ in (\ref{delta12})] in the Hamiltonian  $\PT$-symmetric system (\ref{eq:nonlin1D}).

\begin{figure}
	\begin{center}
		\includegraphics[width=\textwidth]{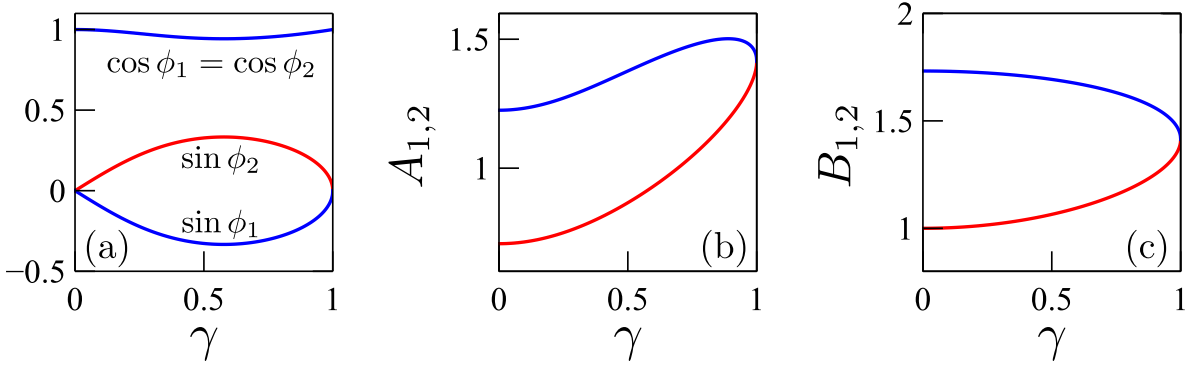}
	\end{center}
	\caption{(a): Dependencies $\cos\phi_{1,2}$ and $\sin\phi_{1,2}$  for $\phi_{1,2}$ in Eq.~(\ref{eq:phi}) for $\Omega=1$ and changing $\gamma$. (b)-(c) Parameters $A$ and $B$ of bright solitons (\ref{eq:AB}) plotted as functions of $\gamma$ for fixed $\Omega=1$ and $\mu=-2$. In all panels, blue and red curves correspond to symmetric and antisymmetric exact bright solitons, respectively [i.e., to upper and lower signs in Eqs. (\ref{delta12})--(\ref{eq:NLS}) and (\ref{eq:solitons})--(\ref{eq:AB})].  Note that $\cos\phi_1=\cos\phi_2$, and the corresponding blue and red curve coincide in (a).}
	\label{fig:AB}
\end{figure}

For the sake of illustration, we plot $\cos\phi_{1,2}$ and $\sin\phi_{1,2}$ in Fig.~\ref{fig:AB}(a).

\subsection{Broken linear $\PT$ symmetry}

For broken linear $\PT$ symmetry ($\gamma \geq \Omega$), we assume that $\sin\delta = -{\Omega}/{\gamma}$, i.e.,
\begin{eqnarray}
\label{delta12EP}
\delta=\delta_1=-\arcsin (\Omega/\gamma) \quad \mbox{or}  \quad  \delta=\delta_2 = \pi +\arcsin(\Omega/\gamma).
\end{eqnarray}
The solution $U\neq0$ and $V\equiv0$ is valid if $\phi=0$,  $\alpha=\pi/4$ and $U$ solves the NLS equation   with linear dissipation ($\delta=\delta_1$) or gain ($\delta=\delta_2$)  
\begin{eqnarray}
\label{NLS_dissip}
 iU_t = -U_{xx} +i\Omega \cot(\delta) U + 2 g|U|^2U.
\end{eqnarray}
The original fields are recovered as 
\begin{equation}
\label{eq:uv}
u=-i\sqrt{2}\sin(\delta/2)U, \quad v=\sqrt{2}\cos(\delta/2)U.
\end{equation}

For $\delta=\delta_1$  or  $\delta=\delta_2$ the total number of particles $N(t)$ decays to zero or grows, respectively. Note that the  decaying solution does not violate   inequality (\ref{eq:N}) since the substitution (\ref{eq:uv})  implies identically zero  quasi-power (\ref{eq:Q}): $Q(t)=Q(0)=0$.  Additionally we note that  solution (\ref{eq:uv}) is asymmetric, i.e., the field amplitudes   $u$ and $v$ are not equal: $|u/v| = |\tan(\delta/2)|$.

In the particular case of the exceptional point (EP) of the underlying linear $\PT$-symmetric system, $\Omega=\gamma$,  Eq.~(\ref{NLS_dissip}) becomes  the integrable conservative NLS equation. Thus the decaying and growing solutions bifurcate from the conservative solution at the EP $\Omega=\gamma$,  with $\gamma/\Omega\geq 1$   being the bifurcation parameter.

\section{Bright solitons and their dynamics}
\label{sec:solitons}

As we have shown in Sec.~\ref{sec:below}, the introduced Hamiltonian $\PT$-symmetric system contains as a particular case the standard NLS equation and therefore  admits various exact solutions. In this section, we explore an important class of solutions in the form of  bright solitons, which can be found for the self-focusing nonlinearity in the NLS equation (\ref{eq:NLS}). We therefore assume $g=-1$. Additionally, in this section we set $\Omega=1$.  Then, using the results of Sec.~\ref{sec:below}, we   readily find two families of bright solitons {\color{black}of system (\ref{eq:nonlin1D})}:
\begin{equation}
\label{eq:solitons}
u_{1,2}(x,t) = e^{-i\delta_{1,2}/2}A_{1,2}\sech(B_{1,2} x) e^{-i\mu t}, \quad v_{1,2}(x,t) = e^{i\delta_{1,2}} u_{1,2},
\end{equation}
where
\begin{equation}
\label{eq:AB}
\eqalign{
A_{1,2} = \frac{\sqrt{2}}{2} (3\gamma^2+1)^{1/4}\sqrt{-\mu \pm \sqrt{1-\gamma^2}}, \quad
B_{1,2} = \sqrt{-\mu \pm \sqrt{1-\gamma^2}}.}
\end{equation}
Subscripts $1$ and $2$ correspond to upper and lower signs in Eqs.~(\ref{delta12})--(\ref{eq:NLS}). In what follows, we call the two identified families   symmetric (with subscript $1$) and antisymmetric (with subscript $2$), because  in the limit $\gamma=0$ the two components are identical for the symmetric solitons ($u_1=v_1$), but are opposite for antisymmetric solitons ($u_2=-v_2$). 

In Eqs.~(\ref{eq:solitons})--(\ref{eq:AB}), $\mu$ is the real parameter which characterizes the temporal frequency of the solution (i.e., the BEC's chemical potential). From Eqs.~(\ref{eq:AB}) it follows that the symmetric solitons exist for $\mu<\sqrt{1-\gamma^2}$, and the antisymmetric solitons require   $\mu<-\sqrt{1-\gamma^2}$. As it is typical for $\PT$-symmetric systems, the solutions constitute a continuous family:   if the parameter $\gamma$ is fixed, one can construct a continuous set of solutions by changing the ``internal'' parameter $\mu$. Notice however, that for $\gamma$ in the interval $(0,1)$ the families of symmetric and antisymmetric solitons do not coexist, because, as readily follows from   Eq.~(\ref{eq:phi}), symmetric solitons require $\phi<0$, and antisymmetric solitons exist for $\phi>0$. For fixed $\mu$,   branches of symmetric and antisymmetric solitons coalesce at $\gamma=1$  as one can observe  in Fig.~\ref{fig:AB}.

\begin{figure}
	\begin{center}
		\includegraphics[width=0.8\textwidth]{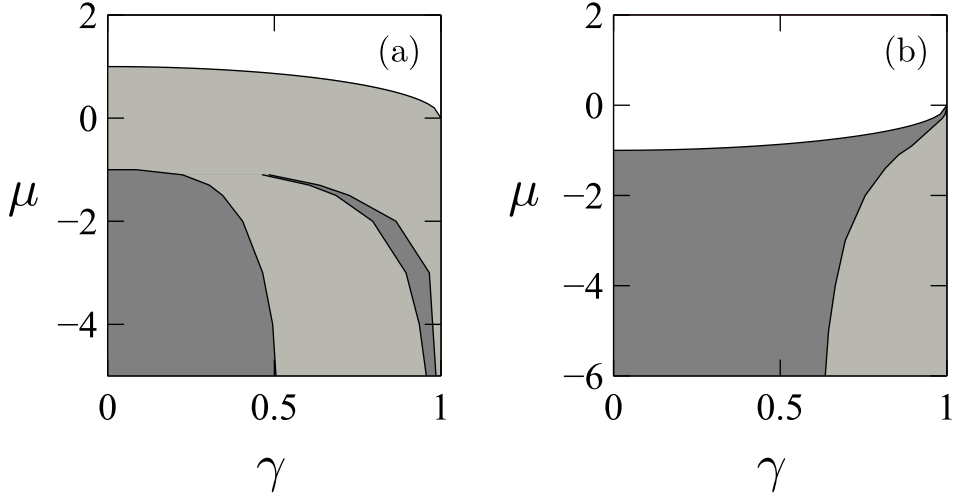}
	\end{center}
	\caption{Stability domains for   bright symmetric (a)  and antisymmetric (b) solitons. White domains correspond to $(\mu, \gamma)$, where the solitons do not exist. Dark gray domains correspond to regions  where the solitons exist and stable; light gray domains correspond to unstable solitons.}
	\label{fig:stab}
\end{figure}

In order to  check   the linear stability of solitons, we used  the substitution 
\begin{equation}
\label{eq:linstab}
\eqalign{
u(x,t) = [e^{-i\delta/2}A\,\sech(B x) + w(x)e^{\lambda t} + z^*(x)e^{\lambda^*t}] e^{-i\mu t},\\%
v(x,t) = [e^{i\delta/2}A\,\sech(B x) + W(x)e^{\lambda t} + Z^*(x)e^{\lambda^*t}] e^{-i\mu t}}
\end{equation}
and   linearized system (\ref{eq:nonlin1D}) with respect to perturbations $(w,  z, W, Z)$. [In (\ref{eq:linstab}) we omitted subscripts $1$ and $2$ because the linear stability substitution  has the same form  for symmetric and antisymmetric solitons.] The growth rates of eventual instabilities,   given by $\textrm{Re}\, \lambda$, were computed numerically from  eigenvalues $\lambda$ of a large sparse matrix which was obtained after approximation of  second spatial derivatives using the second finite difference.  
The outcomes of our stability study are summarized in Fig.~\ref{fig:stab} which shows that solitons of either type are stable if wide parameter ranges.

\begin{figure}
	\begin{center}
		\includegraphics[width=1.00\textwidth]{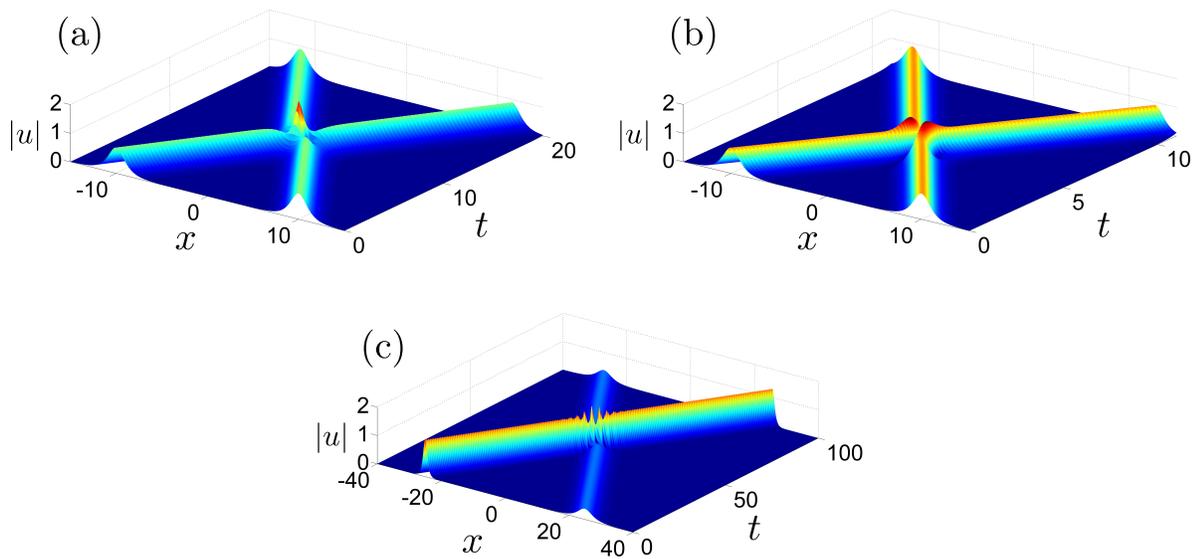}
	\end{center}
	\caption{Interactions of stable antisymmetric solitons. (a) Two identical in-phase solitons with $\mu=-2.5$ and $\gamma=0.2$; (b)  Two identical out-of-phase solitons with $\mu=-2.5$ and $\gamma=0.2$; (c) Two solitons  with  $\gamma=0.3$ and    different amplitudes:  the large-amplitude soliton has $\mu=-4$, and the small-amplitude soliton has $\mu=-0.5$.  All panels show the amplitude of field in the first component, i.e., $|u|$; \textcolor{black}{the behavior of the second component $|v|$ is almost identical to   $|u|$}.}
	\label{fig:dynstab}
\end{figure}

Although the reported  solitons are exact solutions, the system apparently is not integrable. This raises the question about solitons' interactions. \textcolor{black}{In order to address this issue, we prepare the initial condition for system (\ref{eq:nonlin1D}) in the form of a superposition of two   separated solitons (\ref{eq:solitons})  which are launched towards each other:
\begin{eqnarray}
u(x, 0) = u_{j_+}(x-l, 0) e^{-i c x} +  u_{j_-}(x+l, 0) e^{+i c x},\\%
v(x, 0) = v_{j_+}(x-l, 0) e^{-i c x} +  v_{j_-}(x+l, 0) e^{+i c x},
\end{eqnarray}
where constants $l \gg 1$  and $c>0$ determine the initial separation between the solitons and initial solitons' velocity, respectively. Subscripts $j_+$ and $j_-$ can acquire values $1$ or $2$, depending on the type of the soliton (symmetric or antisymmetric). Both solitons correspond to the same value of the gain-and-loss coefficient $\gamma$ [which is the parameter of the system (\ref{eq:nonlin1D})], but, generically speaking, have different chemical potentials $\mu_+$ and $\mu_-$ [because the chemical potential does not enter system (\ref{eq:nonlin1D}) but represents an internal parameter of each soliton]. Once $\gamma$,  $j_\pm$,    and $\mu_\pm$ are chosen, the initial amplitudes and width of the solitons are computed from Eqs.~(\ref{eq:AB}).  	 Then the dynamics of system  (\ref{eq:nonlin1D})  is simulated numerically.}
Quite surprisingly, in our simulations  we observe that  the system behaves as nearly integrable, and solitons interact elastically, similarly to the solitons in the integrable NLS equation. In Fig~\ref{fig:dynstab}(a) we show the collision of two identical antisymmetric in-phase solitons, which pass through each other without any visible distortion. Two out-of phase solitons in Fig.~\ref{fig:dynstab}(b) elastically repel each other  and   recover their original shapes.  Moreover, in spite of the gain and losses, we also  observed    elastic interactions of (stable) solitons with considerably different  amplitudes, as illustrated in Fig.~\ref{fig:dynstab}(c). Similar results were also obtained for collisions of symmetric solitons, when their parameters are chosen from the stability domain.

\begin{figure}
	\begin{center}
		\includegraphics[width=\textwidth]{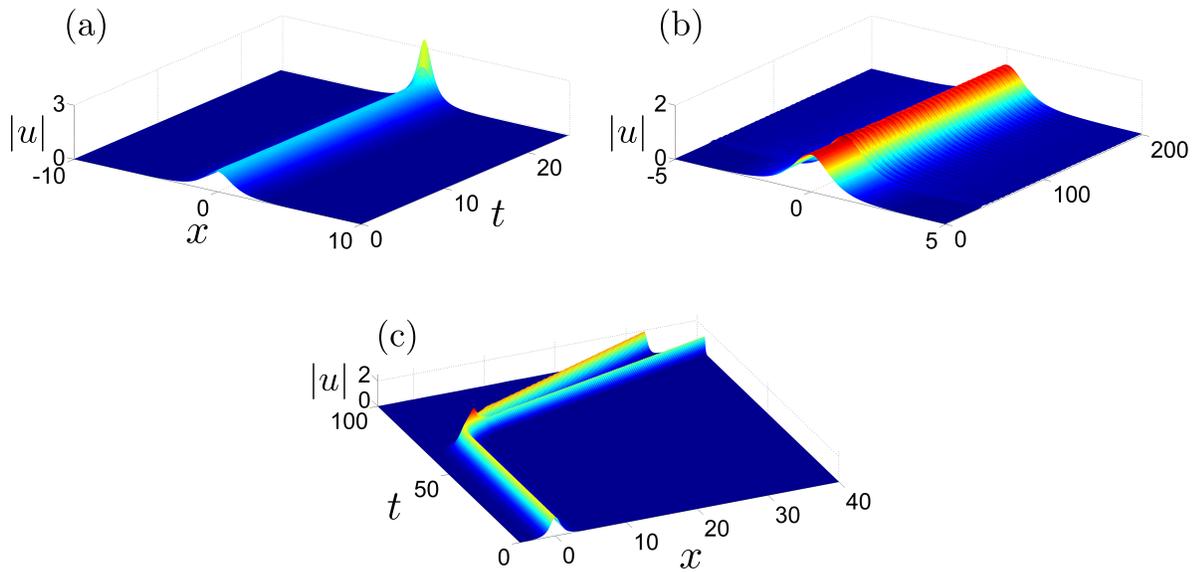}
	\end{center}
	\caption{Different behaviours of unstable solitons. (a) Antisymmetric soliton with $\gamma=0.8$ and $\mu=-2$ exhibits the unbounded growth of $|u|$. (b) Symmetric soliton with $\gamma=0.6$ and $\mu=-2$ evolves into a long-living oscillating state. (c) Symmetric soliton with $\gamma=0.95$ and $\mu=-2$ breaks into a pair of spontaneously moving and long-living solitons. All panels show the amplitude of the field in the first component, i.e., $|u|$; \textcolor{black}{the behavior of the second component $v$ is explained in the text}. }
	\label{fig:dynunstab}
\end{figure}

Finally, we explored  numerically the dynamics of  unstable solitons. \textcolor{black}{ To this end, we numerically integrated system (\ref{eq:nonlin1D}) with initial conditions in the form (\ref{eq:solitons})--(\ref{eq:AB}) with $t=0$ (small-amplitude random distortions have been   introduced in the initial conditions in order to boost the development of an eventual dynamical instability). Three different dynamical scenarios observed for different $\gamma$ and $\mu$} are presented in Fig.~\ref{fig:dynunstab}. The first observed    scenario (which was found the most typical for antisymmetric solitons)  consists in the  infinite growth of the $u$-component (the one that is subjected to the linear gain $i\gamma$) as shown in  Fig.~\ref{fig:dynunstab}(a). \textcolor{black}{Because of  the fast growth of the  amplitude in the first component (i.e., $u$), the numerical process eventually diverges, and the computation terminates; the amplitude of the second   component $v$  remains moderated  and does not grow (at least until the moment when the numerical process diverges).}     For symmetric solitons, we recorded two different scenarios: for relatively small gain-and-loss $\gamma$ the instability  manifests itself in the  emergence of a long-living  oscillating (breather-like) mode [Fig.~\ref{fig:dynunstab}(b)], which is another  pattern typical to  integrable model.  For large gain-and-loss $\gamma$  the initially quiescent 
soliton breaks   into a pair of  pulses which propagate   with different velocities [Fig.~\ref{fig:dynunstab}~(c)].  \textcolor{black}{For dynamics in Fig.~\ref{fig:dynunstab}(b) and (c),   behaviors of   $u$ and $v$  components are almost identical.}

\section{Discussion and Conclusion}
\label{sec:concl}

In this paper, we proposed and investigated a nonlinear $\PT$-symmetric coupler which have several remarkable properties. First, it admits Hamiltonian and Lagrangian formulations, which is not a typical property of nonlinear $\PT$-symmetric systems in general.  Second, it has (at least) three conservation laws which can be derived from Noether's theorem. In the conservative limit, the system becomes bi-Hamiltonian, i.e., admits two different Hamiltonian representations simultaneously. 
Additionally, the introduced $\PT$-symmetric coupler supports a variety of exact solutions which can take form of bright and dark solitons or more complex patterns. Moreover, it was demonstrated numerically that some of the exact solutions are dynamically stable and undergo elastic collisions, similar to collisions of solitons of the integrable NLS    equation. We have also outlined the physical relevance of the introduced model in the context of Bose-Einstein condensates in nonlinear lattices.

While the main goal of this paper was to introduce a dispersive  Hamiltonian $\PT$-symmetric system, the proposed model admits several   generalizations which are worth  future study. The first evident generalization of the Hamiltonian (\ref{eq:Hlin_PT}) suggests to consider the case of two different nonlinear coefficients $g_1$ and $g_2$, i.e.,
\begin{eqnarray}
\label{eq:g12}
H = \int_{-\infty}^\infty \left[ u_x^*  v_x +  u_x    v_x^* + i\gamma (uv^* - u^*v) + \Omega (|u|^2 + |v|^2)  
\right. \nonumber \\ \left. \hspace{2cm}
+  (g_1|u|^2+g_2|v|^2)(e^{i\phi}u^*v + e^{-i\phi}uv^*) \right] dx.
\end{eqnarray}
Then the Hamiltonian equations (\ref{eq:motion}) lead to a generalized version of system (\ref{eq:nonlin1D}):
 \begin{equation}
 \label{eq:nonling12}
 \eqalign{
 	iu_t = -u_{xx} + i\gamma u +\Omega v + g_1 e^{-i\phi}  |u|^2 u + 2g_2 e^{-i\phi}  |v|^2u + g_2e^{i\phi} u^*v^2,\\
 	iv_t = -v_{xx} - i\gamma v +\Omega u + g_2e^{i\phi} |v|^2 v+  2g_1 e^{i\phi} |u|^2v + g_1e^{-i\phi}  u^2v^*.}
 \end{equation}
For $g_1\ne g_2$ this nonlinear system is not $\PT$ symmetric (in the sense discussed in Sec.~\ref{sec:main}). Interestingly enough, this model admits a stationary solitonic solution. Indeed,   under the assumptions
\begin{eqnarray}
u= i\cot^{1/2}(\delta/2)  U, \quad  v= \tan^{1/2}(\delta/2) U,\\
\label{eq:Im2}
\gamma = \Omega\, \textrm {cosec}\, \delta, \quad 
g_2 = -g_1 \cot^2 \delta,
\end{eqnarray}
where   $\delta \in (0, \pi)$ is a  free parameter, system (\ref{eq:nonling12}) reduces to a scalar NLS equation with linear gain or loss (depending on value of $\delta$) and purely imaginary nonlinearity [compare with (\ref{NLS_dissip})]:
\begin{equation}
\label{eq:cGL}
iU_t = -U_{xx} + i\Omega \cot(\delta) U + 2ig_1 \cot(\delta/2) \sin\phi|U|^2U.
\end{equation}
This equation has a well-known stationary solution in the form of the  Pereira-Stenflo soliton \cite{PS77}
\begin{equation}
U = A e^{-i B^2 t}\textrm{sech}^{1+i\sqrt{2}}(B x),
\end{equation}
where 
\begin{equation}
 A^2 = -\frac{3\Omega \cos\delta}{8g_1  \sin\phi(1+\cos\delta)}, \quad B^2 = \frac{\Omega \cot(\delta)}{2\sqrt{2}}.
\end{equation}
The existence of this stationary 
solution is  remarkable in view of  the fact that the coupler operates in the domain of broken $\PT$ symmetry, i.e., $|\gamma/\Omega|\geq 1$ [as readily follows from  the first of equations (\ref{eq:Im2})].

Regarding further generalizations of the introduced models, we notice that while herein we have considered a spatially one-dimensional  dispersive system,  the Hamiltonian structure is expected to survive in the case of multiple dimensions as well, i.e., for $x\in \mathbb{R}^D$, $D\geq 2$. 
Another potential generalization is related to the possibility to address nonlinear multi-component systems with three (or more) coupled waveguides. Some of such dispersive systems should also admit  Hamiltonian structure.

\ack
The research of D.A.Z. is supported by  Russian Science Foundation (Grant No. 17-11-01004).  
V.V.K. was supported by the FCT (Portugal) Grant No. UID/FIS/00618/2013.

\end{document}